\documentclass[12pt,preprint]{aastex}

\newcommand{\etal}{{\em et al.}}            

\shorttitle{Complex X-ray Morphology of NGC 7618}
\shortauthors{Kraft \etal}

\begin{document}

\title{The complex X-ray morphology of NGC 7618:  A major group-group merger in the
local Universe?}
\author{R. P. Kraft}
\affil{Harvard/Smithsonian Center for Astrophysics, 60 Garden St., MS-67, Cambridge, MA 02138}
\author{C. Jones}
\affil{Harvard/Smithsonian Center for Astrophysics, 60 Garden St., MS-2, Cambridge, MA 02138}
\author{P. E. J. Nulsen}
\affil{Harvard/Smithsonian Center for Astrophysics, 60 Garden St., MS-6, Cambridge, MA 02138}
\author{M. J. Hardcastle}
\affil{University of Hertfordshire, School of Physics, Astronomy, and Mathematics, Hatfield AL10 9AB, UK}

\begin{abstract}

We present results from a short {\em Chandra}/ACIS-S
observation of NGC 7618, the dominant central galaxy of
a nearby ($z$=0.017309, d=74.1 Mpc) group.
We detect a sharp surface brightness discontinuity
14.4 kpc N of the nucleus subtending an angle of 130$^\circ$
with an X-ray tail extending $\sim$ 70 kpc in the opposite direction.
The temperature of the gas inside and outside the discontinuity
is 0.79$\pm$0.03 and 0.81$\pm$0.07 keV, respectively.
There is marginal evidence for a discontinuous change in the elemental
abundance ($Z_{inner}$=0.65$\pm$0.25,$Z_{outer}$=0.17$\pm$0.21 at
90\% confidence), suggesting that this may be an `abundance' front.
Fitting a two-temperature model to the ASCA/GIS spectrum of the
NGC 7618/UGC 12491 pair shows the
presence of a second, much hotter ($T$=$\sim$2.3 keV) component.
We consider several scenarios for the origin of the edge and the tail
including a radio lobe/IGM interaction, non-hydrostatic `sloshing',
equal-mass merger and collision, and ram-pressure stripping.  In the last
case, we consider the possibility that NGC 7618 is falling into UGC 12491,
or that both groups are falling into a gas poor cluster potential.
There are significant problems with
the first two models, however, and we conclude that
the discontinuity and tail are most likely the result of ram
pressure stripping of the NGC 7618 group as it falls into a
larger dark matter potential.

\end{abstract}


\section{Introduction}

The complex cluster morphology seen in X-ray images and in galaxy distributions gave support
to the hypothesis that large structures form hierarchically; that is, that small
groups of galaxies merge to form low-mass subclusters, which then merge to form a massive
rich cluster with the infalling groups aligned along large filaments.  
Galaxies and groups are the building blocks of the observable Universe and
contain the bulk of the observable baryons.  Groups are estimated to
contain a significant fraction, 20-30\%, of the total matter in the Universe.
Thus groups are important cosmological indicators of the distribution and properties of
the dark matter.  However, because they are not as luminous as clusters,
they have received less study than their more massive cousins.

Observations of rich clusters show many examples of pending or ongoing
mergers of subclusters.  In particular in X-ray cluster catalogs, about
40\% of rich clusters show substructure \citep{jon84, jon99, moh95}.
Virtually all stages of cluster mergers
have been thoroughly investigated with both the {\em Chandra} and XMM-Newton
observatories \citep{mar00,vik01,bri04,hen04}.
However, less attention has been paid to the merging of
groups and the formation of low-mass clusters due to their lower X-ray luminosity and
the paucity of examples in the local Universe.
To our knowledge, the only nearby example of the early stages of the merger of two roughly equal
mass groups is the NGC 499/NGC 507 pair \citep{kim95,kra03}.
In the hierarchical scenario, group mergers
represent a critical transitional phase in the formation of larger scale structure.
An understanding of the group merger process is thus fundamental to understanding
the growth of structure.

In this paper, we report results from analysis of a short {\em Chandra}/ACIS-S observation
of the nearby elliptical galaxy NGC 7618 ($z$=0.017309 or d$_L$=74.1 Mpc
for WMAP cosmology \citep{spe03} - 1$''$=350 pc).  The X-ray luminosity of NGC 7618 is
$\sim$7$\times$10$^{42}$ ergs s$^{-1}$ in the 0.1-10 keV band, typical of groups, not
isolated elliptical galaxies, although it appears to be optically
isolated \citep{col01}.
We find a sharp surface brightness discontinuity in the X-ray emission north of 
of the nucleus of NGC 7618, and an extended tail to the south.  
We conclude that these features are either the result of a
major group-group merger with UGC 12491, a group
that lies 14.1$'$ on the sky from NGC 7618 at virtually identical
redshift ($z$=0.017365) \citep{ebl02}, or ram-pressure stripping due
to infall of NGC 7618 into a larger gravitational potential (which may include
UGC 12491).  
This pair has been poorly studied in large part because of its relatively low galactic
latitude ($\ell=105.575$, $b=-16.909$, $A_V$=0.97, $N_H$=1.19$\times$10$^{21}$ cm$^{-2}$)).

\section{{\em Chandra} and ASCA Observations}

NGC 7618 was observed for 18.4 ks with Chandra/ACIS-S on
December 10, 1999 (OBSID 802).  The lightcurve of events in the 5.0-10.0 keV
bandpass on the entire S3 chip, excluding the NGC 7618 nucleus and any point sources
visible by eye, was created using 259 s bins and examined for periods of flaring background.
Intervals where the rate was more than 3$\sigma$ above the mean rate were removed.
There was considerable background flaring during this observations, and
almost 10 ks of data were excluded.  Only 8438 s of good time remained.  Bad pixels,
hot columns, and columns along node boundaries were also removed.
We present data from the S2 and S3 chips in this paper.
Absorption by foreground gas in our galaxy ($N_H$=1.19$\times$10$^{21}$ cm$^{-2}$) was
included in all spectral fits.
NGC 7618 was also observed by ASCA for $\sim$54 ks on July 7, 1998.
UGC 12491 is contained within the FOV of the ASCA/GIS, and we use
these data to measure the temperature of the gas around this galaxy
and the diffuse emission between NGC 7618 and UGC 12941.
Data from the SIS was not used because of its smaller field of view.

\section{Analysis}

\begin{figure}
\plotone{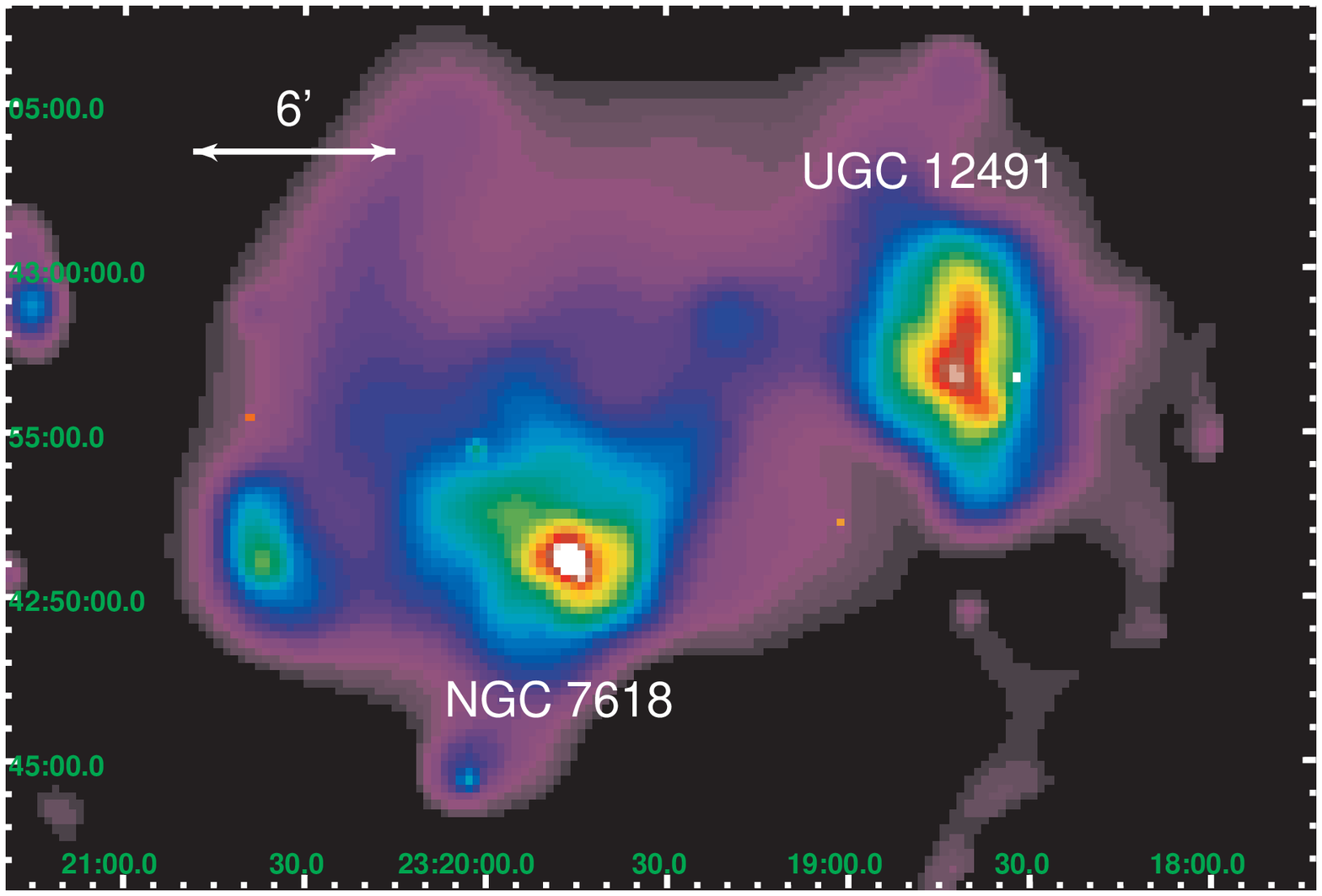}
\caption{Adaptively smoothed ASCA/GIS image of NGC 7618 and
UGC 12491.  The minimum (violet) and maximum (white) surface
brightnesses in this image are $\sim$38.4 and 134.2 cts arcmin$^{-2}$, respectively.
The background, estimated from
an off-source region at the edge of the field of view is
23.5$\pm$1.1 cts arcmin$^{-2}$.  The violet regions
in this image are therefore significant at more than 10$\sigma$.}\label{gisimg}
\end{figure}

An adaptively smoothed ASCA/GIS image of the NGC 7618/UGC 12491 pair
(usin the CIAO program `csmooth') is shown in Figure~\ref{gisimg}.
The optical and X-ray properties of both groups are summarized in
Table~\ref{galtab}.
Their X-ray luminosities are each $\sim$6-7$\times$10$^{42}$ ergs s$^{-1}$, and their
X-ray emission extends to radii of 150-200 kpc. 
The luminosities and spatial extents far exceed those of
typical isolated elliptical galaxies and are more representative of
poor clusters or fossil groups \citep{vik99}.
The central elliptical galaxies represent only a small fraction of
the gravitating mass that resides in a much larger dark matter halo.
No optical census of the galaxy populations around either NGC 7618 or UGC 12491
has been undertaken, but their recessional velocities differ by only 17 km s$^{-1}$.
The temperature of the gas around these galaxies is $\sim$0.8 keV,
typical of groups.  
Based on their X-ray luminosities, X-ray extents, and spatial proximity we
conclude that these two objects are groups likely within $\sim$300 kpc of each other
and gravitationally interacting.

\begin{figure}
\plotone{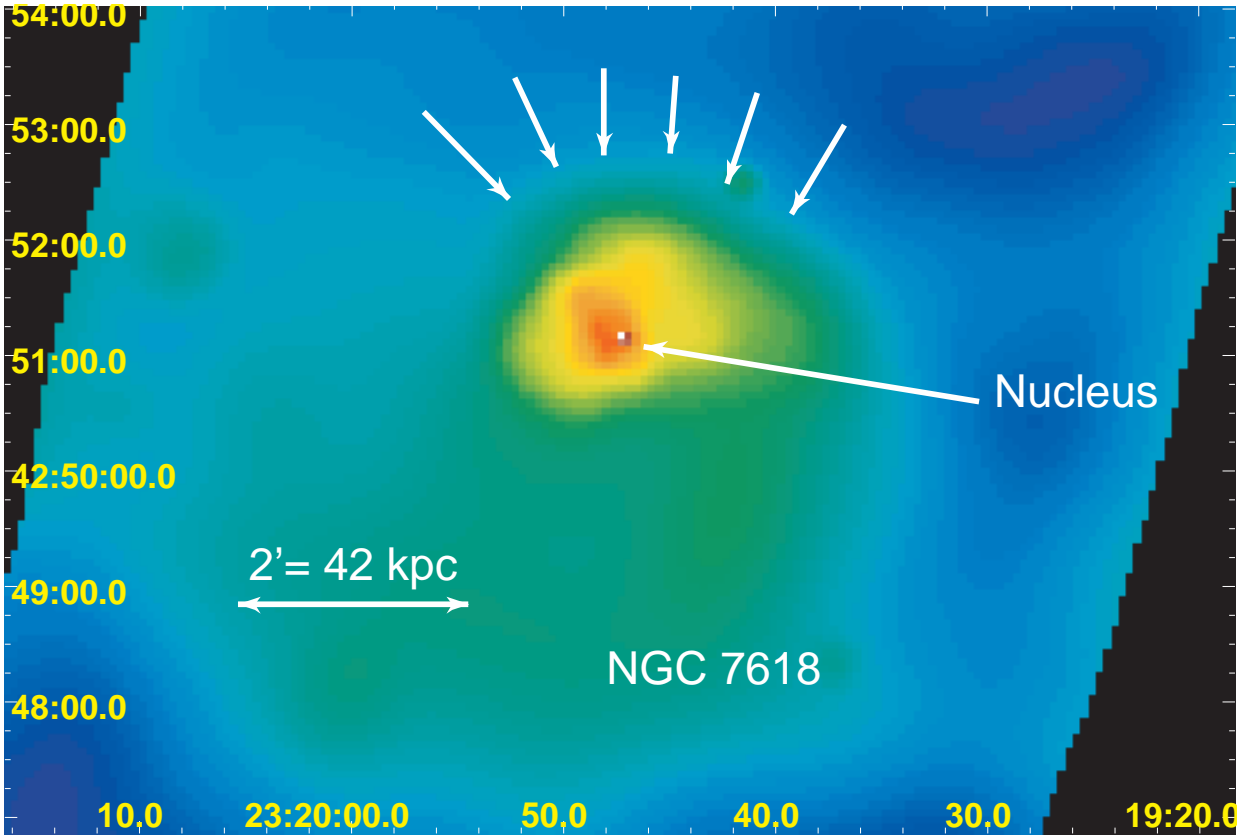}
\caption{Adaptively smoothed, exposure corrected, background subtracted
Chandra/ACIS-S image of NGC 7618 in the 0.5-2.0 keV band.  The
white arrows denote the surface brightness discontinuity.}\label{acisimgb}
\end{figure}

\begin{figure}
\plotone{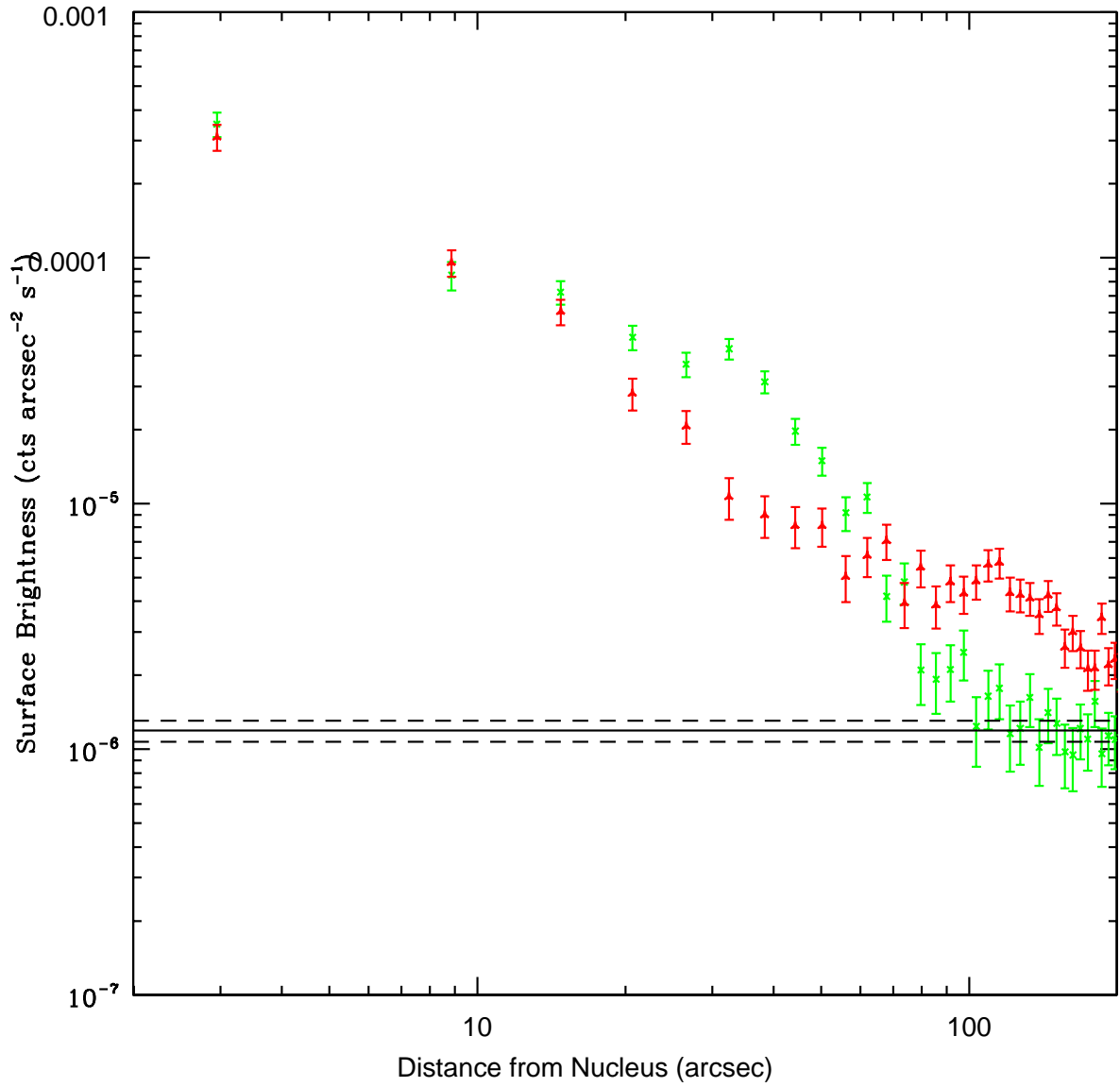}
\caption{Surface brightness profiles in 90$^\circ$ sectors to the North (green)
and South (red) centered on the nucleus of NGC 7618.  The solid and dashed lines
indicate the background level and uncertainty, respectively, estimated from
a distance region of the S3 chip.  Note the excess of counts in the Southern `tail'
from 70$''$ to 200$''$.}\label{tailwedge}
\end{figure}

\begin{figure}
\plotone{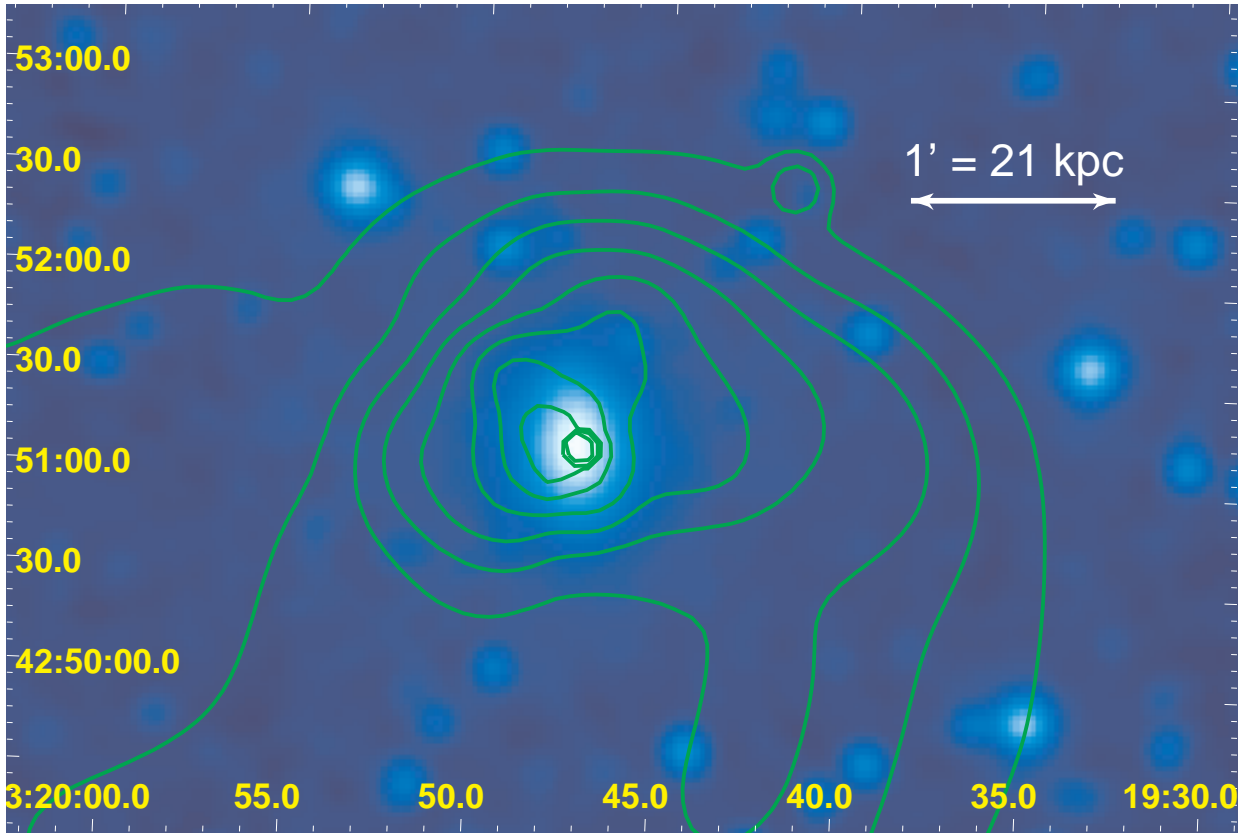}
\caption{Contours from previous image overlaid onto
optical/DSS image of NGC 7618.  The active nucleus is
co-incident with the peak of the optical isophotes.  The contours correspond
to a surface brightness of 3, 5, 9, 16, 28, 50, 88, 155, 274, and 484
$\times$10$^{-5}$ cts arcsec$^{-2}$ s${-1}$.}\label{xoptovl}
\end{figure}

An adaptively smoothed, exposure corrected, background subtracted
{\em Chandra}/ACIS-S image in the 0.5-2.0 keV band
of NGC 7618 is shown in Figure~\ref{acisimgb}.
The X-ray bright active nucleus of the host galaxy is labeled at the center.
There are three unusual features to note in this image.
First, the peak of the diffuse X-ray emission (shown orange in
Figure~\ref{acisimgb}), and therefore presumably
the peak in gas density, as well as the center of the host
galaxy lie $\sim$1$'$ North of the center of the larger scale diffuse
X-ray emission (shown green).
An X-ray `tail' extends $\sim$3.3$'$ (69.3 kpc) South of the nucleus.
The statistical significance of this `tail' can be seen in the
X-ray surface brightness profiles shown in Figure~\ref{tailwedge}.
These profiles were made in two 90$^\circ$ sectors
to the North (green) and South (red) with the nucleus of NGC 7618 at
the vertex.  The surface brightness in the S sector at distances larger than 70$''$ from
the nucleus is several times higher than in the North sector.
Either the dark matter has a very unusual distribution, or the gas is
not in hydrostatic equilibrium in the gravitational potential.
Second, there is a sharp surface brightness discontinuity $\sim$41$''$ (14.4 kpc)
North of the nucleus that spans $\sim$130$^\circ$.  This discontinuity
is delineated by the white arrows in Figure~\ref{acisimgb}.
and is probably a contact discontinuity between two moving fluids.
Third, on smaller scales,
the X-ray emission from the gas peaks $\sim$9$''$ (3.2 kpc) to the E of the
nucleus of the host galaxy as seen in Figure~\ref{xoptovl}.
In the central 10 kpc of the host galaxy, the stars will dominate the gravitating
mass, so there clearly is an offset between the hot gas and the gravitating mass.
All of these strongly suggest that the gas has partially
separated from the gravitating matter and indicate non-hydrostatic motions.

There is an X-ray point source coincident with the optical nucleus, presumably
a low-luminosity AGN.  The point source contains 48 counts in the 0.5-5.0 keV band.
Assuming a power-law spectrum with photon index 1.7 and galactic
absorption, the X-ray luminosity of the active nucleus is
4.2$\times$10$^{40}$ ergs s$^{-1}$ in the 0.1-10.0 keV band (unabsorbed).
The X-ray luminosity of the nucleus is roughly an order of magnitude larger than
that expected based on the correlation of X-ray and radio cores \citep{can99}.
This may be a considerable underestimate if the nucleus is heavily absorbed,
but is typical of that found in other ``normal'' elliptical galaxies \citep{jon05}.

\begin{figure}
\plotone{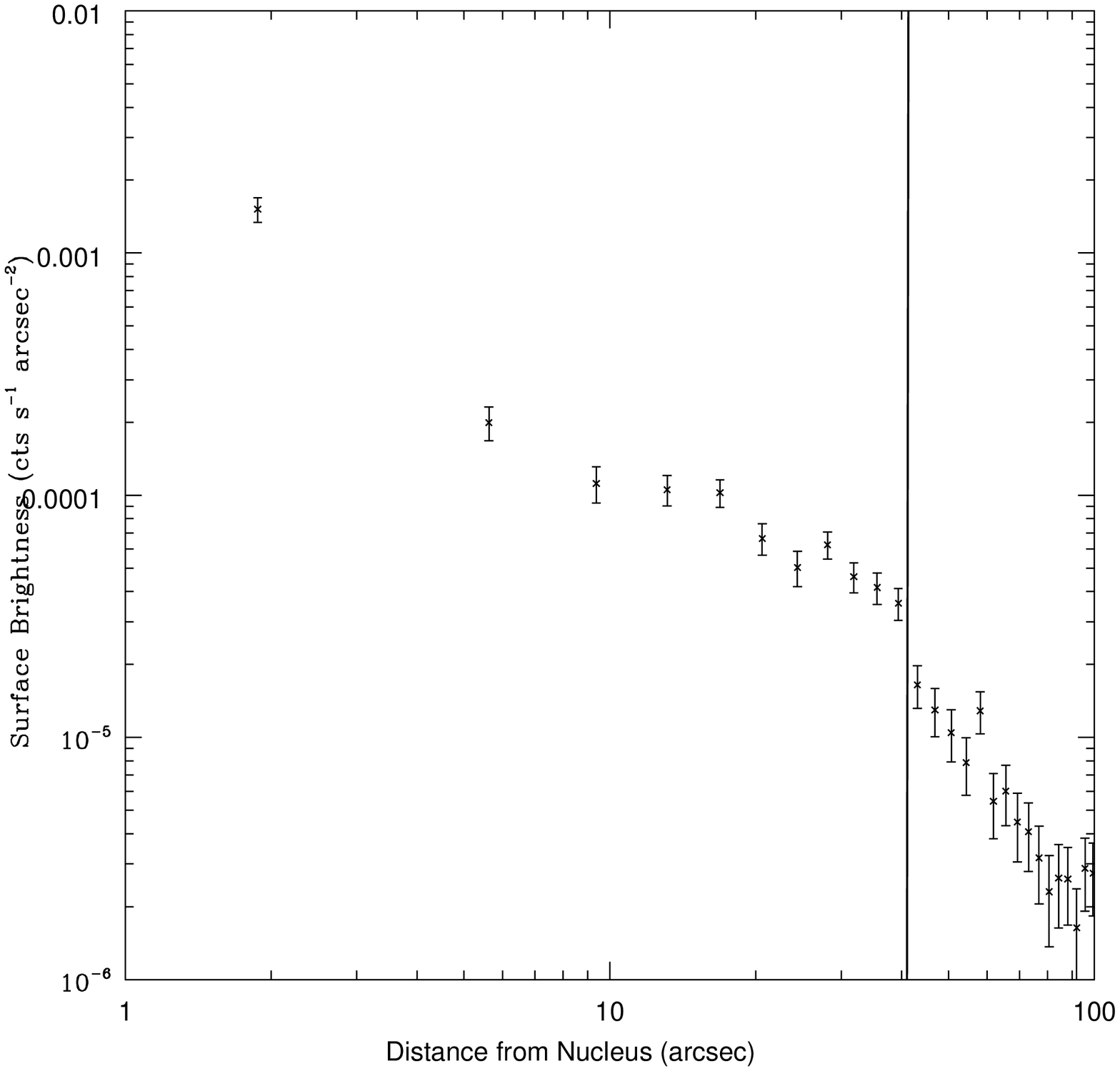}
\caption{Surface brightness profile in a 60$^\circ$ sector N
of the nucleus across the discontinuity.  The position of the surface
brightness discontinuity is denoted by the dark vertical line.}\label{sbprof}
\end{figure}

\begin{figure}
\plotone{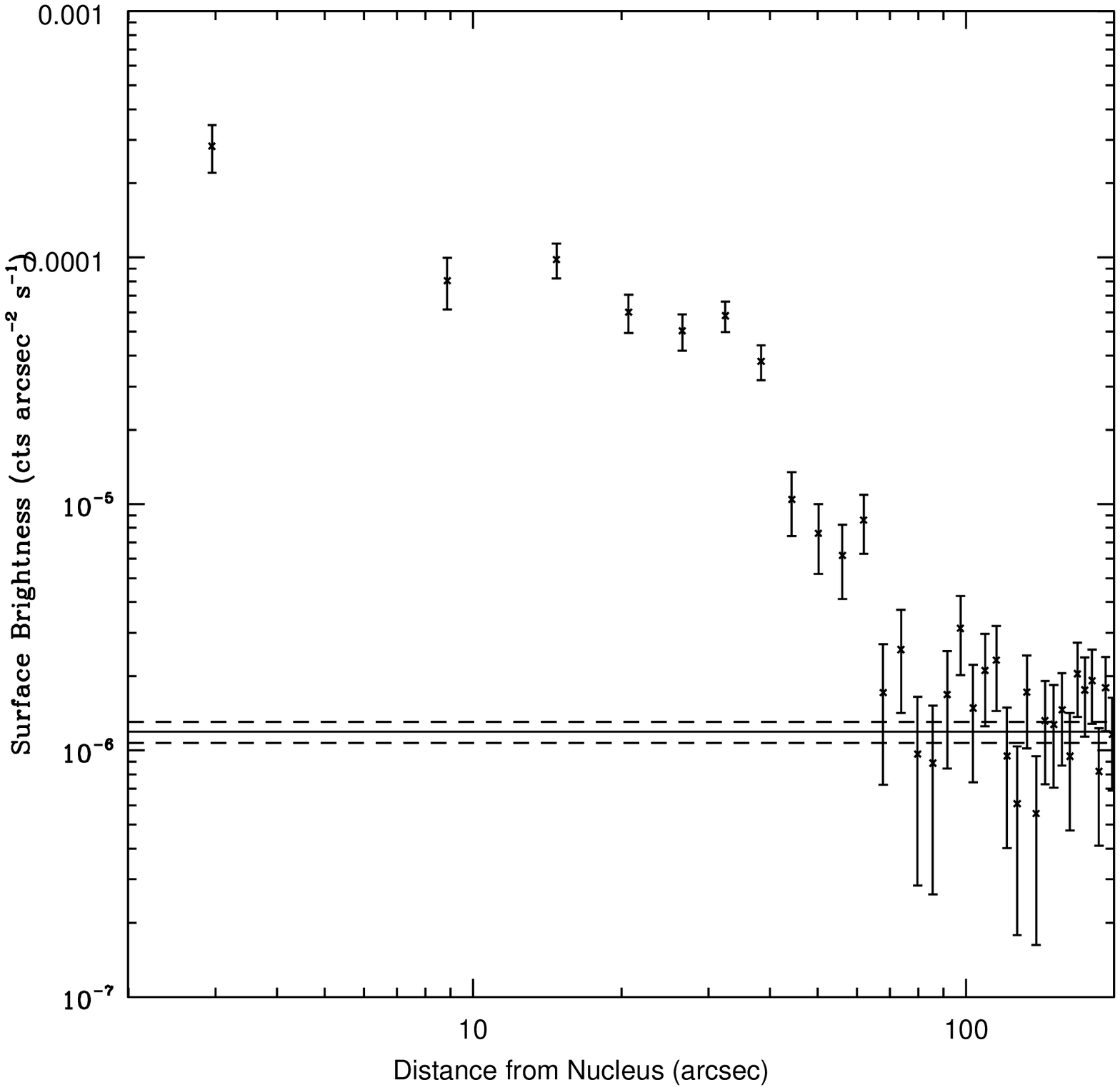}
\caption{Surface brightness profile in a 30$^\circ$ sector to the NE 
of the nucleus across the discontinuity.  The position of the surface
brightness discontinuity is denoted by the dark vertical line.}\label{newedge}
\end{figure}

The surface brightness profile in a 60$^\circ$
sector to the North of NGC 7618, shown in Figure~\ref{sbprof},
drops by approximately a factor of two
across the discontinuity.  At the sharpest region of the edge to the NE of
the nucleus, the surface brightness drops by a factor of 
4 over a distance of $\sim$4$''$ (1.4 kpc).
The surface brightness profile in a 30$^\circ$ sector to the NE is shown
in Figure~\ref{newedge}.
The morphology of this discontinuity is similar to that seen in {\em Chandra} observations
of cluster 'cold-fronts', although there is no evidence for a temperature discontinuity
between the two moving fluids as commonly seen in clusters of galaxies \citep{vik01,mar01,maz01}.
We fitted absorbed APEC models to the spectra in 90$^\circ$ sectors of two
annular regions, one inside the discontinuity and one outside.
The thickness of the inner and outer annuli were 33.0$''$ (11.6 kpc) 
and 66.4$''$ (23.2 kpc), respectively.
The gas temperature does not change significantly
across the discontinuity ($T_{inner}$=0.785$\pm$0.025 keV,
$T_{outer}$=0.810$\pm$0.070 keV). 
There is marginal evidence for a jump
in the elemental abundance ($Z_{inner}$=0.65$\pm$0.25, $Z_{outer}$=0.17$\pm$0.21 - all
uncertainties at 90\% confidence), however.
This suggests that the discontinuity could be an `abundance' front due to
a sharp discontinuity in the elemental abundance, and therefore the emissivity
of the gas as observed in NGC 507 \citep{kra03}.
The temperature and elemental abundance structure in the
gas is probably more complex than we have assumed here, but the short
exposure time and limited quality of the data prevent a more detailed analysis.

Fitting power laws to the surface brightness profiles interior to and
exterior to the discontinuity and assuming hydrostatic equilibrium, we find
a large change in the power law index across the discontinuity,
$\beta_{in}$=0.30 and $\beta_{out}$=0.64.
The change in $\beta$ is almost certainly not related to a change
in the gravitational potential and suggestive of non-hydrostatic gas motions.
We estimate the gas density on both sides of the discontinuity by
deprojecting the surface brightness profile (assuming spherical symmetry) and find 
proton densities of 6.0$^{+1.2}_{-0.8}$ and 5.0$^{+2.4}_{-1.2}$
$\times$10$^{-3}$ inside and outside the discontinuity, respectively.
The upper limit of the velocity of the gas interior to the discontinuity estimated
from the maximum pressure difference is Mach 0.9 or $\sim$ 420 km s$^{-1}$ \citep{vik01}.

The X-ray morphology of the ASCA/GIS image (Figure~\ref{gisimg}) suggests that
the two groups reside in a larger-scale dark matter potential.
A third X-ray peak is seen 8.5$'$ to the east of NGC 7618, and diffuse emission
extends at least 15$'$ to the north and 10$'$ to the south.
Any gas that resides in this larger scale dark matter halo should
be hotter than the gas in the NGC 7618 group.  We fitted the ASCA/GIS spectrum
in three regions, two 7$'$ radius circles centered on each galaxy and
a third region 15$'$ in radius centered between NGC 7618 and UGC 12491 excluding
the two 7$'$ radius circles centered on each galaxy.
The radius of the two circles centered on NGC 7618 and UGC 12491 corresponds
to the 90\% encircled energy radius for the ASCA/GIS in the 1-2 keV band.
Background was determined from ASCA/GIS high-latitude, blank sky
observations taken from the HEASARC and generated using the FTOOL `mkgisbgd'.

We fit absorbed, single temperature APEC models to each spectrum with $N_H$ frozen at
the Galactic value and the elemental abundance frozen at 0.6$Z_\odot$.
The results of these fits are summarized in the top half of Table~\ref{spectab}.
Single temperature models are poor fits for the regions centered on
NGC 7618 and UGC 12491, but provide
an adequate description of the diffuse emission between the galaxies.
The temperature of the diffuse gas between NGC 7618 and UGC 12491
is 2.32$^{+0.50}_{-0.34}$.
We also fit two temperature models to the two spectra extracted from
the regions centered on NGC 7618 and UGC 12491.  As before, both the $N_H$ and the elemental
abundance were frozen.  For NGC 7618, we also froze the temperature 
of one component at 0.8 keV,
the value determined from the ACIS-S spectral fitting, to reduce the
number of free parameters.  If this parameter
is allowed to freely vary, the fit value is consistent with 0.8 keV.  For UGC 12491,
the temperatures of both components were allowed to freely vary.
The results of these fits are summarized in the bottom half of Table~\ref{spectab}.
The two temperature model provides acceptable fits in both cases, and
the temperature of the hotter component is $\sim$2.3 keV.

\section{Discussion}

There are at least four possible explanations for the observed X-ray
structures.  First, the complex X-ray morphology could be the result
of a radio lobe/IGM interaction.  
NGC 7618 is a radio source. It is detected in the NVSS with a flux
density of 20 mJy, with some evidence of extension NW/SE.
At higher frequencies (5 and 8 GHz) two 15-min archival
VLA observations only detect a point-like core coincident with the
center of the galaxy, with a flux density of 4.5 mJy at 4.9 GHz and
3.0 mJy at 8.4 GHz. However, at lower frequencies, the flux density
is much higher: 1.2 Jy in the 151-MHz 6C catalog \citep{hal93},
0.26 Jy in the 408-MHz B3 catalog \citep{fic85},
and 1.0 Jy in the 74-MHz VLA Low-Frequency Sky
Survey (VLSS: http://lwa.nrl.navy.mil/VLSS/). The high flux density
at low frequencies suggests that there may be a relic radio source,
the aged synchrotron plasma from a more energetic phase of the
active nucleus.
The X-ray morphology is very different from that seen in other examples
of radio lobe/ISM interactions, however.
There are no obvious X-ray cavities as are commonly seen in such radio 
lobe/ISM interactions (e.g. \citep{mcn00,jon02,hei02}),
nor is there evidence of hot, shock-heated shell
that would be present if the radio lobes were expanding supersonically \citep{kra03}.
It is not clear whether any of these
features would be expected in the case of a relic radio source, however.
A sensitive low-frequency radio map would give us a better
understanding of the possible interaction between radio-emitting
plasma and the hot gas.

Second, it is possible that the NGC 7618 gas is oscillating, or 'sloshing', in the gravitational 
potential because of a recent merger/interaction with a lower mass sub-group.
A dust lane has been detected in the optical host galaxy, perhaps
indicative of a recent merger and adding support to this hypothesis \citep{col01}.
Similar structures (on larger scales) to those presented
here have been seen in X-ray observations of clusters of galaxies \citep{mar01,tit05}.
In this scenario, the gas is oscillating, or `sloshing', in the dark matter
potential due to a recent merger with a sub-cluster.  The infall of the sub-cluster
drives a shock into the gas that displaces it from the gravitating matter.  The
gas oscillates around the center of mass, eventually returning to hydrostatic
equilibrium via viscous dissipation.
Such processes can contribute a significant amount of
energy to the gas and may disrupt or prevent the formation of cooling flows in clusters of
galaxies.  If we naively assume the gas on both sides of the
discontinuity is in hydrostatic equilibrium, we find
an unphysical discontinuity in the gravitating mass.
The apparent mass discontinuity is the result of the 
non-zero acceleration of the gas inside the discontinuity.
The distribution of the gas will reflect the reduced gravity.
That is, the acceleration term in Euler's equation is non-zero for gas
inside the discontinuity.
Assuming that the core is at maximum displacement from the center, the gravitational potential
energy of the gas is $U\sim G\Delta M r^{-1} \sim$7$\times$10$^{14}$ ergs gm$^{-1}$,
or roughly one third of the thermal energy of the gas.

There is one important difficulty with this interpretation, however.  On tens of
kpc scales, the emission peak of the gas and the optical galaxy (NGC 7618) are both
offset (to the N) relative to the larger
scale X-ray isophotes (shown green in Figure~\ref{acisimgb}).  
If the sloshing scenario is correct only the gas
and not the galaxy/gravitating mass, should be moving relative to the large scale dark
matter halo.  The host galaxy should be resting at the center of the gas
distribution.  In addition, the peak of the X-ray emission lies to the East
of the nucleus of the host galaxy, but the larger scale `sloshing' is North/South.  
This suggests that there are significant gas motions along perpendicular axes, and
that the gas is coupled to the gravitating mass (the stars) 
along the North/South axis, but partially decoupled along the East/West axis.
It is difficult to see how there could be the case unless the merger were off-axis.  If this is the case,
the merging galaxy should be detectable by an optical census.

The third possibility is that the observed structures
are the result of a 'near-miss' flyby of the nearby UGC 12491 group.
In this scenario, the UGC 12491 group passed by NGC 7618 from the SE toward 
its current position to the WNW, and
the complex X-ray morphology of NGC 7618
is the result of fluid motions in the gas induced by the
gravitational and hydrodynamical interaction.  
The observed X-ray morphology is the result of the NGC 7618 core 
being displaced relative to the larger scale gravitational potential, and  
the surface brightness discontinuity represents a discontinuity in the elemental abundance.  
X-ray observations of groups and clusters show that the elemental abundance
is strongly peaked toward the center.  The displacement of the core of a group
relative to its halo could create the observed structures.

Hydrodynamic simulations of head-on collisions of equal mass clusters
show that the close approach of the cores can create strong non-hydrostatic motions in 
the gas \citep{rot97}.
Elongation of the dark matter potential, and thus the gas distribution, is a natural
consequence of these collisions.
In these simulations, the gas becomes partially separated from the gravitating matter and
is not a good tracer of the dark matter
distribution.  The separation between UGC 12491 and NGC 7618 was
probably (much) smaller in the past in order
to create such a large disturbance in the NGC 7618 gas.
The simulations of merging clusters show little separation of the gas from
the dark matter as the two clusters initially approach each other.
It is only as the cores collide/merge and subsequently separate
that significant gas velocities develop.
If the current positions of the two groups
are, in fact, their closest approach so far, non-hydrostatic
effects should just be starting to manifest themselves.
Therefore, the two groups must have already passed each other at least once.

There are several difficulties with this scenario, however.
If the hotter gas seen in the ASCA/GIS spectrum is shock-heated group gas,
the measured temperature ratio (0.8 to 2.3 keV) implies
an infall/approach velocity of the two groups of $\sim$1170 km s$^{-1}$ \citep{lan89},
much larger than typical peculiar velocities.
It would also be difficult to explain the large amount of hot gas ($\sim$10$^{12}$ M$_\odot$)
seen in the ASCA/GIS observation in this scenario under the assumption that
this component is entirely shock-heated group gas.
The total (thermal) energy of the hotter component is only $\sim$1.3$\times$10$^{61}$ ergs,
a reasonable value for the collison of two groups (we note that an AGN
outburst could easily supply this much energy to the gas as well \citep{mcn04}).
If the hot component is, in fact, shock-heated group gas, it is
not bound to the group potential and a transient phenomenon, escaping as
a wind.

It is likely that these two groups are gravitationally bound.
Ignoring angular momentum (which is not believed to be important
on large scales \citep{hof86}) and dissipative forces, and assuming that
the mass of each group is 10$^{13}$ M$_\odot$, the escape velocity of this
pair is $\sim$550 km s$^{-1}$.  If either of these groups has a velocity
relative to the center of mass larger than this value, the pair will be
unbound.  This is larger than typical peculiar velocities of galaxies/groups,
and it is reasonable to conclude that the system is bound.
Significant angular momentum will decrease this value, whereas dissipative
forces will increase it.  The dynamic parameters of this pair are too
poorly known to make a definitive statement, however.

The fourth possibility is that NGC 7618 is falling into
a larger gravitational potential.  In this scenario, either
UGC 12491 is at the center of the potential and NGC 7618 is
falling into it, or both NGC 7618 and UGC 12491 are falling
into a larger scale potential.
The existense of the hotter ($\sim$2.3 keV)
gas component in the ASCA/GIS observation (Figure~\ref{gisimg})
supports this scenario.
The X-ray structures seen in the gas around NGC 7618 are then
the result of ram-pressure stripping.
This is the classic `cold-front' scenario discussed by \citet{vik01}.
The effects of ram-pressure stripping on the hot gas atmospheres of
early-type galaxies falling into clusters have been studied in {\em Chandra}
observations of NGC 4472, NGC 4552 (falling into the Virgo cluster) and NGC 1404
(falling into/toward NGC 1399, the dominant member of the Fornax cluster) \citep{bil04,mac05a,mac05b}.  
In these cases, a sharp surface brightness discontinuity 
in the direction of infall and a diffuse tail in the opposite direction
have been observed.  The gas around NGC 7618 is being stripped by
hotter, lower density gas that presumably resides in the larger/deeper dark
matter halo.  

If this scenario is correct, either NGC 7618 is falling into UGC 12491 or
both groups are merging with/falling into a larger dark matter potential in
which there is no central dominant elliptical galaxy.
The high temperature of the second spectral component in the ASCA analysis and the
lack of an azimuthally symmetric X-ray halo around UGC 12491 (Figure~\ref{gisimg})
support the latter hypothesis.
This phenomenon has in fact already been observed on a smaller
scale in the Pegasus I group \citep{kra05}.  Neither
of the massive ellipticals in this group, NGC 7619 or NGC 7626, lie at the
center of the extended X-ray halo.
{\em Chandra} observations of NGC 7619 show a sharp surface brightness
discontinuity to the NE (presumably the direction of infall), and an
extended, ram-pressure stripped tail in the opposite direction.

If we assume the hotter (2.3 keV) component is in hydrostatic equilibrium
with the dark matter potential and that the gas density follows
a beta-model profile with $\beta$=0.67 and $r_0$=50 kpc (typical for clusters
of galaxies),
the gravitating mass, $M_{grav}$, within a radius of 325 kpc of the midpoint between NGC 7618 
and UGC 12491 is $\sim$5.6$\times$10$^{14}$ M$_\odot$.
The gas mass, $M_{gas}$, required to account for the observed ASCA/GIS flux of
the hotter component within this
radius is $\sim$7$\times$10$^{11}$ M$_\odot$ for this density profile.
Ignoring the stellar component, which is insignificant on these spatial scales,
the baryon fraction, $f=M_{gas}/M_{grav}$ is $\sim$1.5\%.
Such a low value of baryon fraction is not improbable for a 2.3 keV
cluster.  We speculate that this larger dark matter halo may be a `failed' cluster
in which much of the gas was blown off during a cataclysmic event
early in its formation (either a merger or a powerful AGN remnant).  
We caution, however, that the spatial resolution of the ASCA/GIS is poor and
our knowledge about the morphology of the gas limited.  In addition, this gas may
not be in hydrostatic equilibrium, so the uncertainties on both (gas and
gravitating) mass estimates are large.  A moderate XMM-Newton observation of this pair could measure
the temperature and morphology of the hotter component and resolve this issue.

We note that in this scenario the surface brightness discontinuity north of the NGC 7618 nucleus
cannot be the stagnation point between the hot gas in NGC 7618 and the
gas of this putative larger scale halo.
If the observed X-ray surface brightness discontinuity
is in fact the stagnation point between
these two gases, the temperature of the gas exterior to the discontinuity must be
on the order or higher than that of the halo gas (from Bernoulli's equation).
The stagnation point therefore must lie beyond the observable emission, and the
surface brightness discontinuity represents a contact discontinuity between two fluids.
Unless the infall is highly supersonic, the gas interior to the stagnation point should
not be highly disturbed and should remain in rough hydrostatic equilibrium in the
gravitational potential of NGC 7618.
A deeper {\em Chandra} observation of the central regions of
NGC 7618 is required to elucidate the hydrodynamics of the gas.

\section{Conclusions}

We have observed a sharp surface brightness discontinuity in the X-ray emission
from the hot gas in the NGC 7618 group, and an X-ray `tail' extending
70 kpc in the opposite direction in an 8 ks {\em Chandra}/ACIS-S observation.
Archival ASCA/GIS observations indicate the presence of a hotter (2.3 keV)
component, although the morphology of this gas is poorly constrained.
We conclude that there are three possible explanations for these features.
First, the NGC 7618/UGC 12491 pair underwent a recent `near-miss' flyby.
If this is the case, this pair is the nearest early-stage merger of two
roughly equal mass groups.
Second, UGC 12491 may be at the center of a cluster, and NGC 7618 is falling into
it.  Third, NGC 7618 and UGC 12491 are both falling into a gas poor cluster with
no dominant central elliptical galaxy.
Whether the observed features are the result of a group-group merger,
or the infall of two groups into a larger dark matter potential, the NGC 7618/UGC 12491
pair is one of the best examples of an ongoing merger in the local Universe.
Deeper X-ray observations are required to better constrain the thermodynamic
parameters of the gas in the central regions and the larger scale halo.  Radio
observations will be critical in assessing the role of radio plasma/IGM interactions.
If the observed X-ray features are the result of ram-pressure stripping or a merger
interaction between the groups, the effects
on the relic radio halo are likely to have been dramatic.
A detailed optical census including velocities of the other galaxies in the both groups will
be useful to constrain their dynamics and their relationship to
the larger dark matter potential.

\acknowledgements

This work was supported by NASA contracts NAS8-38248, NAS8-39073,
the Chandra X-ray Center, and the Smithsonian Institution.
We would like to thank the anonymous referee for comments
that improved this paper.

\clearpage

\clearpage

\clearpage

\clearpage

\clearpage

\clearpage

\clearpage

\begin{table}
{\small
\begin{center}
\begin{tabular}{|l|c|c|}\hline
      &  NGC 7618  & UGC 12491 \\ \hline
$m_B$ & 14.0  & 14.9 \\ \hline
$M_B$ & -21.3 & -20.2 \\ \hline
$z$  & 0.017309 & 0.017365 \\ \hline
Distance (Mpc) & 74.1 & 74.3 \\ \hline
$L_X$ (ergs s$^{-1}$) & 6.9$\times$10$^{42}$ & 6.2$\times$10$^{42}$ \\ \hline
X-ray Radius & $\sim$170 kpc & $\sim$200 kpc \\ \hline
\end{tabular}
\caption{Summary of the X-ray and optical properties of the
NGC 7618 and UGC 12491 galaxies.  
The X-ray luminosity is in the 0.1-10 keV
bandpass (unabsorbed) within 7$'$ (146 kpc) of the nucleus.  Absolute magnitudes have
been corrected for extinction.}\label{galtab}
\end{center}
}
\end{table}

\clearpage

\begin{table}
{\small
\begin{center}
\begin{tabular}{|l|c|c|c|}\hline
     &  NGC 7618  &  UGC 12491  &  Diffuse \\ \hline\hline
\multicolumn{4}{|c|}{Single Temperature Fits} \\ \hline
$k_BT$ (keV)  & 1.43$^{+0.08}_{-0.14}$ & 1.32$^{+0.10}_{-0.13}$ & 2.32$^{+0.50}_{-0.34}$ \\ \hline
Flux & 3.06$\times$10$^{-12}$ & 5.37$\times$10$^{-12}$ & 2.53$\times$10$^{-12}$ \\ \hline
$\chi^2_\nu$ & 1.95   &  1.60  & 0.73 \\ \hline\hline 
\multicolumn{4}{|c|}{Two Temperature Fits} \\ \hline
$k_BT_1$ (keV)  & 0.80 & 0.86$^{+0.16}_{-0.09}$ &  \\ \hline
Flux & 2.26$\times$10$^{-12}$ & 3.69$\times$10$^{-12}$ &  \\ \hline
$k_BT_2$ (keV)  & 2.21$^{+0.33}_{-0.67}$ & 2.26$^{+1.19}_{-0.35}$ & \\ \hline
Flux & 1.48$\times$10$^{-12}$ & 2.15$\times$10$^{-12}$ & \\ \hline
$\chi^2_\nu$ & 1.03   &  0.78  & \\ \hline\hline 
\end{tabular}
\caption{Best-Fit Temperatures and Fluxes of the ASCA/GIS data in three regions.
All uncertainties are 90\% confidence for one parameter of interest.  Units of
fluxes are ergs cm$^{-2}$ s$^{-1}$ (unabsorbed) in the 0.5-2.0 keV band.}\label{spectab}
\end{center}
}
\end{table}

\end{document}